\documentclass[twocolumn,prl,superscriptaddress,showpacs]{revtex4}%
\usepackage{graphicx}
\usepackage{amsmath}
\usepackage{amssymb}
\usepackage[ansinew]{inputenc}
\newcommand{\figwidth}{\columnwidth}

\begin{document}
\title{Dissipation-driven quantum phase transitions in a Tomonaga-Luttinger liquid \\
electrostatically coupled to a metallic gate}
\author{M.~A. Cazalilla}
\affiliation{Donostia International Physics Center (DIPC), Manuel de Lardizabal 4,
20018-Donostia, Spain,}
\author{F. Sols} 
\affiliation{Departamento de F\'{\i}sica de Materiales, Universidad Complutense, Madrid, E-28040, Madrid, Spain,}
\author{F. Guinea}
\affiliation{Instituto de Ciencia de Materiales de Madrid, CSIC, Cantoblanco, E-28043 Madrid, Spain.}
\begin{abstract}
  The dissipation induced by a  metallic gate on the low-energy properties of 
 interacting 1D electron liquids is studied. As function of the distance
 to the gate, or the electron density in the wire,
the system undergoes a quantum phase transition  from the Tomonaga-Luttinger
 liquid state to  two kinds of dissipative phases, one of them with a finite spatial
 correlation length. We also define a dual
 model, which describes an attractive one dimensional metal with a Josephson
 coupling to a dirty metallic lead.
\end{abstract}
\maketitle

\emph{Introduction:}  In the study of  properties of quantum wires 
(and other mesoscopic systems),  proximity to metallic gates is  
frequently regarded as a  source of static screening of the interactions in the wire. 
Within a classical electrostatic picture, which is valid at long distances
from the gate, this arises because  electrons in the wire interact, not only 
amongst themselves, but also with their image charges in the
gate.  Thus the Coulomb potential becomes a 
more rapidly decaying dipole-dipole potential. Moreover, metallic 
environments (\emph{e.g.} gates) are also a source of de-phasing and 
dissipation~\cite{CB01,GJS04,G05,MB05} (for experimental research on related
topics, see~\cite{WPG05}), which arise because the electrons in the wire
can exchange energy and momentum with 
the low-energy electromagnetic modes of the gate. 
This effect is described by the dissipative part of the
screened potential and, as we show below, it can lead to backscattering (\emph{i.e.}
a scattering process where one or several electrons reverse their direction of motion) 
in a one-dimensional (1D) quantum wire. We find that, for an arbitrarily small coupling to the gate,  the backscattering can drive a quantum phase transition  provided  that the
 interactions between the electrons in the wire are sufficiently repulsive.
\begin{figure}
\includegraphics[width=0.9 \figwidth]{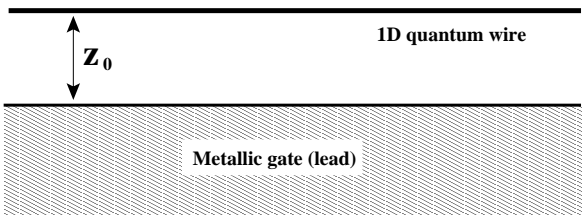}
\caption{1D quantum wire
 near a metallic gate or spin-gapped quantum wire near a dirty  metallic lead.\label{fig1}}
\end{figure}

The effects of dissipation on quantum phase transitions have attracted much 
attention~\cite{W91,KMKC01,RDOF03,LLH05}. We discuss here transitions induced
by dissipation, as in~\cite{WTS05,WT04,WVTC05}.
Some aspects of this work are also related to previous research on
1D systems coupled to 
environments~\cite{CCN97,N97,G02,G03,MG05,HD06}. In particular, 
the system studied here can be considered a 
microscopic realization of the model studied in~\cite{WTS05}. 
A discussion of the relations between our findings and this
work will be given below.

\emph{The model:} The electrostatic coupling of the wire to a metallic gate 
is described by
\begin{equation}
H_{gw} = \int d{\bf r} d{\bf r}'\, v_c({\bf r} - {\bf r}') \rho_{w}({\bf r}) 
\rho_{g}({\bf r}'), \label{eq1}
\end{equation}
where $v_c({\bf r}) = e^2/ 4\pi \epsilon |{\bf r}|$ is the (statically screened)
Coulomb potential, where $e < 0$ is the electron charge and $\epsilon$ the
dielectric constant of the insulating medium located between  the gate
and the wire. The operators $\rho_g({\bf r})$ and $\rho_{w}({\bf r})$ describe
the  density fluctuations in  the gate and the wire, respectively.
For a 1D quantum wire, $\rho_w({\bf r}) \simeq \rho_w(x) \delta(y) 
\delta(z -z_0)$, where
$z_0 \gg w >0$ is the distance measured from the surface of  the gate (see
Fig.~\ref{fig1}) and $w$ is the width of the wire. Following~\cite{G84,SG87}, 
we integrate out the density modes of the  metallic gate, 
and obtain the following effective action for the 1D wire 
($L$ is the length of the wire and $T = \beta^{-1}$ is the temperature):
\begin{eqnarray}
S_{\rm diss} &=& \frac{1}{2\hbar} \int^{L}_0 dx dx' \int^{\hbar\beta}_0 d\tau d\tau'
\: \rho_w(x,\tau) \nonumber\\
&&\times V_{\rm scr}(x-x',z_0,\tau-\tau')  \rho_w(x',\tau'), \label{eq2}
\end{eqnarray}
where we have assumed a translationally invariant flat gate so that the screened
interaction $V_{\rm scr}({\bf r},{\bf r}',\tau) = - \langle V({\bf r},\tau) 
V({\bf r}',0) \rangle_g/\hbar$ , depends only on
${\bf R} - {\bf R}'$ (${\bf r} = ({\bf R},z) = (x,y,z)$);
 $V({\bf r},\tau) = \int d{\bf r}'\, v_c({\bf r} - {\bf r}') \rho_g({\bf r}')$ and $\langle \ldots \rangle_g$ denotes average over the gate's degrees of freedom.
To obtain  an effective description of the effect of (\ref{eq2}) on the low-temperature and 
low-frequency  properties of the 1D wire, we  first employ
bosonization~\cite{H81,GNT98,G04}. In the absence of the gate,
the wire~\footnote{We first consider the simpler case of
spinless fermions and discuss further below how to extend our results to other
systems such as spin-$\frac{1}{2}$ electrons and nanotubes.} 
is a Tomonaga-Luttinger liquid, described by the  action~\cite{H81,G04,GNT98}:
\begin{equation}
S_0[\phi] = \frac{1}{2\pi g} \int dx d\tau \left[ \frac{1}{v} 
\left( \partial_{\tau} \phi \right)^2
 + v \left( \partial_x \phi \right)^2\right], \label{eq4}
\end{equation}
where $\phi(x,\tau)$ is a plasmon field~\footnote{By construction\cite{H81}  
$\phi(x,\tau)$ is an angle defined modulo $\pi$.} 
that varies slowly on the  scale of lattice constant, $a$, and  $v = v_F/g$ is the phase velocity of  the plasmons, being $v_F$ the  \emph{bare} Fermi velocity.
Treating the ions in the wire as a positive uniform background of the 
same   density $\rho_{0}$ as the electrons, we  focus on the density
fluctuations about $\rho_0$. In bosonized form,
$\rho_{w}(x,\tau) = \frac{1}{\pi} \partial_x\phi(x,\tau) +  \frac{1}{2\pi a} \sum_{m \neq 0} 
e^{2i m \left(k_F x  + \phi(x,\tau) \right) }$.  
Each of the terms describes the low-energy 
density fluctuations of momentum $q \approx 2m k_F$,  
where $m$ is an integer and $k_F$  the Fermi momentum.
Replacing $\rho_w(x,\tau) \to \frac{1}{\pi} \partial_x \phi(x,\tau)$
in~(\ref{eq2}) yields a term that describes forward ($q\approx 0$) scattering between
electrons. Such a term leads to the (static) screening 
of the Coulomb interaction  described in the introduction, and its  dissipative
part yields a contribution to the action of the form $\int dq d\omega \,  
|\omega| q^2 f(q) \: |\phi(q,\omega)|^2$. We find
at small $q$ that $f(q) \sim  \ln(1/q)$ for a semi-infinite 3D diffusive gate or
a granular gate. Thus, the forward scattering term is  
irrelevant in the renormalization-group (RG) sense. For a 2D diffusive gate gate, 
$f(q) \sim q^{-1}$ and the
term is marginal in the RG sense but it can be shown~\cite{CCN97,unpub} 
that it does not modify the  power-law correlations  of  the system 
at zero-temperature. Moreover, in the limit of strongly repulsive
interactions which interests us here its effect is  small.
We therefore focus on the 
(backscattering) terms  with $q \approx 2 k_F$, which
describe Friedel oscillations. In the limit of very
repulsive  interactions,  the electrons in 1D wire are a Tomonaga-Luttinger liquid
(TLL) that is close to a Wigner-crystal 
state~\cite{G04,GNT98} and density correlations are dominated 
by this oscillating  term.
Thus, 
\begin{eqnarray}
S_{\rm diss}[\phi]  = -\frac{\eta}{\pi} \int^{L}_0 dx \int^{\hbar\beta}_0  d\tau d\tau' \, 
{\cal K} (\tau-\tau') \nonumber\\
\times  \cos 2  p \left[\phi(x,\tau) - \phi(x,\tau')\right],
\label{eq3}
 \end{eqnarray}
where $\eta \propto  a^{-2} S(Q_p,z_0)$,  $S(q,z_0)$ 
being the Fourier transform  of the  spatial dependence of the 
dissipative part of  $V_{\rm scr}(x,z_0,\tau)  
\simeq W_{\rm scr}(x,z_0)  \delta(\tau) +  {\cal K}(\tau) S(x,z_0)$  
at low energies~\cite{G84,SG87}.  
The static part, $W_{\rm scr}(x,z_0)$, 
has the  effect of screening  the interactions in the wire,
and therefore leads to an effective dependence of $g$ on  $z_0$: 
as $z_0 \to 0$, the more screened the Coulomb interactions are, and therefore  
$g\to 1$. The dissipative kernel   ${\cal K} (\tau) = (\pi/\hbar \beta)^{1+s} 
\left| \sin(\pi \tau/\hbar\beta) \right|^{-1-s}$, for $\tau \gg \tau_c$, 
and $\tau_c =  \omega^{-1}_c$, where $\omega_c = \min\{E_F/\hbar,\omega^{G}_{c}\}$
$E_F$ being the bandwidth of the wire and $\omega^{G}_{c}$ 
the characteristic response frequency of the gate electrons~\footnote{The dissipative part is $\propto \omega$ to leading order. The first correction is $A\:  \omega^3$,
where $A$ depends strongly on details of the gate response function and on $z_0$. 
To be able to compare it to the leading term, we use dimensional analysis and write
$A = (\omega^{G}_{c})^{-2} {\cal A}$, where $\omega^{G}_c \approx \hbar v^G_F/\ell_p$
for a diffusive gate with mean free path $\ell_p$ and Fermi velocity   $v^G_F$.
For a clean gate $\omega_c \approx \min\{\omega_p, E_F/\hbar\}$ being $\omega_p$
the plasma frequency (3D gate) and $E^G_F$ the Fermi energy}.  We have
generalized the model to consider general dissipative environments
(characterized by $s$,  a metallic gate corresponding to 
\emph{ohmic} dissipation, $s = 1$) 
as well as generic backscattering processes for 
$q\approx Q_p = 2 p^2 k_F$.
Spinless electrons correspond to $p=1$, spin-$\frac{1}{2}$ electrons
to $p = \sqrt{2}$ (provided that $g < \frac{1}{3}$~\cite{G04}), 
and nanotubes to $p = 2$ 
(provided that $g < \frac{1}{5}$~\cite{GCFGB04}).

The dependence of the friction coefficient $\eta$  on the 
gate-wire distance $z_0$  can be obtained for various models of the gate: 
$\eta \approx a^{-2} (\pi Q_p \sigma_{2D})^{-1} L_0(2 Q_p z_0)$ for a 2D gate, 
and  $\eta \propto  a^{-2} (\pi \sigma_{3D})^{-1}  K_0(2 Q_p z_0)$ 
where $\sigma_{\rm 2D}$ and $\sigma_{\rm 3D}$ are gate conductivities 
measured in units of $e^2 / \hbar$,  and $L'_0(x) = K_0(x)$, $K_0(x)$
being the modified Bessel function of the 2nd kind. 
Thus, the values  of $\eta$ and $g$ can be tuned either by bringing the wire 
closer to the gate or, alternatively if the wire is connected to leads, by 
charging the gate to vary the  chemical potential 
(and therefore the density $\rho_0 \propto  k_F \propto Q_p$) of the wire. In deriving Eq.~(\ref{eq3}), 
we have assumed that the wire is away from half-filling. The analysis 
of the half-filled case is more involved and will reported elsewhere~\cite{unpub}.

\emph{Weak coupling RG analysis:} To assess the stability of the TLL
when perturbed by $S_{\rm diss}[\phi]$ we have  studied
the RG flow of the above model.  Assuming that the 
dimensionless coupling  $\alpha =  ( v \tau_c)(\tau_c)^{1-s}   \eta$   is small,
we  perturbatively  integrate high-frequency density fluctuations 
to lowest order in $\alpha$, and  
obtain the following RG equations:
\begin{eqnarray}
\frac{dv}{d\ell} &=&  -4 p^2 g v\: \alpha , \quad \frac{dg}{d\ell} 
=  -4 p^2 g^2 \: \alpha \label{eq5}\\
\frac{d \alpha}{d\ell} &=& (2 - s -2 p^2 g)\: \alpha, \label{eq6}
\end{eqnarray}
where  $\ell = \ln (\omega_c/T)$.
These equations describe a Kosterlitz-Thouless-like transition around a
quantum critical point where $\alpha = \alpha^* =0$ and  
$g = g^* = (2-s)/2p^2$  ($g^{*} = 1/2p^2 $ for the ohmic case). 
At the critical point correlations decay as 
power-laws with universal exponents determined by $g^*$.  
\emph{E.g.} density correlations at $2k_F$
decay as $(x^2 + v^2 \tau^2)^{-g^*}$, which implies that the dynamical exponent
$z = 1$. It also interesting to point out that dissipation does not drive any
phase transition for $s > 2$. Furthermore, for an infinite-range dissipative kernel
($s = -1$) and $p = 1$,  (\ref{eq5}, \ref{eq6}) reduce to the equations
derived by Voit and Schulz~\cite{VS87} and Giamarchi and Schulz~\cite{GS88} 
for  spinless electrons in the presence of phonons and disorder, respectively. 

 In the phase where $\alpha$ flows towards strong coupling,  the system is 
characterized by a  length  scale which diverges as $\xi_1 \approx k^{-1}_F
\: e^{-\pi/[(2-s)p\sqrt{\alpha -\alpha_c}]}$  with the distance $\alpha -
\alpha_c$ to the transition,  and behaves as  
$\xi_1 \approx k^{-1}_F  \: \left[\alpha (0)\right]^{\frac{1}{2p^2(g-g^*)}}$ 
far from it. On the side where $\alpha$ scales down to zero, the system is a TLL 
with infinite conductivity at zero
temperature (frequency). However,  at finite temperature (frequency)
the (optical) conductivity is finite a behaves as a power-law: 
$\sigma(T) \sim \frac{1}{\alpha} \: T^{2-\mu}$ (${\rm Re} \, \sigma(\omega > 0) 
\sim \alpha \: \omega^{\mu -4}$, and ${\rm Re} \, 
\sigma(\omega > 0)  \sim \alpha/(\omega \ln^2(\omega))$ at the transition), 
where the exponent $\mu = 2p^2 g + s + 1$. In the phase
where $S_{\rm diss}[\phi]$ is relevant, an illustrative but rather 
crude estimate of the conductivity,  hopefully valid in the large $\eta$
limit, can be obtained by expanding the cosine in Eq.~(\ref{eq3}) 
and keeping the quadratic terms  in $\phi(x,\tau)$ only. 
Using the Kubo formula~\cite{G04}, 
\begin{equation}
\sigma(\omega) = \frac{i {\cal D}}{\omega + i/\tau_d}\label{eq7}
\end{equation}
where ${\cal D} = g e^2 v /\hbar \pi$  is the Drude weight
and  $\tau^{-1}_d = 4 \pi p^2 g  v \eta$. However, for $\omega \gg \tau^{-1}_d$ 
(but  $\omega \ll \omega_c$), we expect a crossover to a power-law  such 
that ${\rm Re} \: \sigma(\omega) \sim \alpha\,  \omega^{\mu-4}$.

\emph{Self-consistent Harmonic Approximation} (SCHA):
A better approximation than expanding the cosine in~(\ref{eq3}) 
can be obtained by using the SCHA, which approximates $S_{\rm diss}$  
by a quadratic term~\cite{S87,DAR03,G04,COG05,HD06}   
$- \frac{1}{2}\int dx d\tau d\tau' \,  \Sigma(\tau-\tau') 
\left[ \phi(x,\tau) - \phi(x,\tau')  \right]^2$. Assuming
that $\Sigma(\tau) \simeq \tilde{\eta}/(\pi \tau^{2})$ at long times  ($s = 1$), 
and optimizing the free energy, we find that $\tilde{\eta} \sim
(v\tau_c)^{-1}\:  [\eta (v\tau_c)]^{\frac{1}{1 - 2p^2 g}}$. Note that 
$\tilde{\eta} \to 0$  as $g \to g^{*} = \frac{1}{2p^2}$ thus signaling a transition 
and in agreement with the previous RG analysis. Away from the transition 
($g < g^*$), the SCHA yields a  diffusive plasmon propagator,  
$G^{-1}(q,\omega) \simeq \tilde{\eta} |\omega|  + v q^2/(\pi g)$ 
at small $\omega$, signaling the breakdown of the TLL. Moreover,  
$\Phi(x,\tau) = \langle e^{2ip \phi(x,\tau)} e^{-2ip\phi(0,0)} 
\rangle_{\rm SCHA} \simeq   N^2_0 [1 +  C_0(\tilde{\eta} x,\tilde{\eta} v\tau/g)]$, where 
$N_0  = N_0(\tilde{\eta} v \tau_c)$ is non-universal and 
$C_0(\tilde{\eta} x,0) \sim (\tilde{\eta} |x|)^{-1}$ 
whilst $C_0(0,\tilde{\eta} v \tau/g) \sim ( \tilde{\eta} | v\tau/g|)^{-1/2}$.

\emph{Large $N$ approach:} Further insight into the properties of the model
can be obtained by means of the large $N$ approach, which captures many of 
the properties of dissipative quantum rotor models~\cite{SCS95,R97,DAR03,PFGKS04}. 
Let  ${\bf n}(x,\tau) =  \left(\cos 2 p\phi(x,\tau), \sin 2 p \phi(x,\tau) \right)$, so that 
the  action $S = S_0 + S_{\rm diss}$ becomes:
\begin{eqnarray}
S[{\bf n}, \lambda] &=& \frac{1}{2} \int \frac{dq d\omega}{(2\pi)^2} \:
 G^{-1}_0(q,\omega) |{\bf n}(q,\omega) |^2 \nonumber \\
 &+& \frac{i}{2} \int dx d\tau \lambda(x,\tau) 
\left[ {\bf n}^2(x,\tau) - 1 \right],
\end{eqnarray}
where    $G^{-1}_0(q,\omega) = \eta |\omega| +  \kappa_p [ (\omega/v)^2 + q^2]$,
$\kappa_p = 4 p^2 v/(\pi g)$,  and $\lambda(x,\tau)$ is a Lagrange multiplier 
ensuring that ${\bf n}^2(x,\tau) = 1$.  After generalizing
the symmetry  of the model from $O(2)$ to $O(N)$,  
the field ${\bf n}$ is integrated out. In the large $N$ limit, 
the path-integral is dominated by a saddle point 
at $\lambda(x,\tau) =    -i \kappa_p \xi^{-2}$, which obeys:
\begin{equation}
N \int \frac{dq d\omega}{(2\pi)^2} \, G(q,\omega|\lambda = -i \kappa_p \xi^{-2}) = 1,
\label{normalization}
\end{equation}
where $G^{-1}(q,\omega; \lambda_0) = G^{-1}_0(q,\omega) + i \lambda_0$.
For a single ohmic quantum rotor, only solutions of
(\ref{normalization}) with $\xi > 0$ exist~\cite{R97,DAR03}. However,
the spatial coupling of the rotors described by the $q^2$ term of the action,
allows for solutions with $\xi = 0$.  Thus we find  $\xi^{-1} \sim 
(\eta_c-\eta)$, which implies that the critical exponent $\nu = 1$ and $z = 2$, 
in the large $N$ limit~\cite{PFGKS04} (more accurate estimates, $\nu = 0.689(6)$ 
and $z = 1.97(3)$,  have been reported in~\cite{WTS05}). Critical correlations 
 $\Phi(x,\tau) = \langle {\bf n}(x,\tau) \cdot {\bf n}(0,0) \rangle_{N \to \infty} = 
N C_0(\eta x, v \eta v \tau/g)$, with $C_0(x,\tau)$ defined as above 
(a more correct  form for $N = 2$ can be found  in~\cite{WTS05}). 
Away from criticality,  $\Phi(x,0) \sim e^{-|x|/\xi}/| \xi^{-1} x|$ and $\Phi(0,\tau) \sim 
\xi^2 g^2/(v \tau)^{2}$ in the phase with $\xi \neq 0$. For $\eta > \eta_c$, 
we set $n_1(x,\tau) = N_0$ and integrate out the remaining $N-1$
components of ${\bf n}$, and proceed as above. Assuming that the 
phase is ordered,\emph{i.e.} $N_0 \neq 0$, implies that $\xi^{-1} = 0$. 
The correlation function $\Phi(x,\tau)$ takes the same asymptotic form  
as that found using the SCHA.
\begin{figure}
\includegraphics[width=\figwidth]{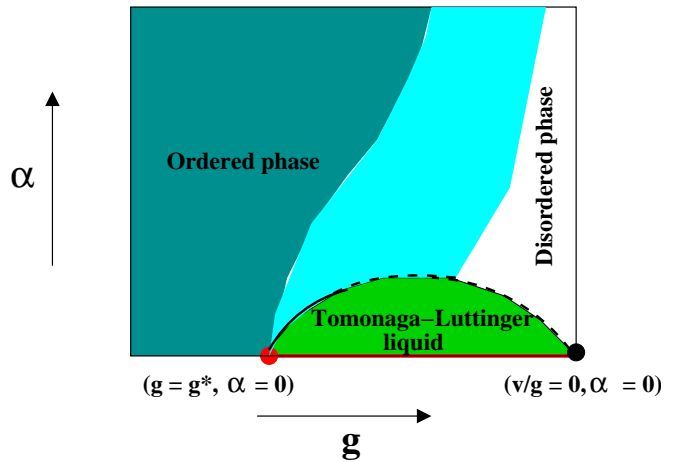}
\caption{Schematic  phase diagram. The 
  quantum critical point at $(g = g^*, \alpha=0)$
  is predicted by the weak coupling RG approach 
discussed in the text.  The dashed lines as well as the phase 
boundary between the ordered and the disordered
phase (lightly shaded region) cannot be 
inferred by the  methods used in this work. 
\label{phase_diagram}}
\end{figure}

 \emph{ Phase diagram:} the simplest flow and phase diagram compatible with
 all above results is shown in Fig.~\ref{phase_diagram}. We find three phases: i)
The TLL phase, where the coupling to the gate flows towards
zero, ii) An ordered, gapless, dissipative phase, which has diffusive plasmons, 
and iii) A disordered phase with a finite spatial correlation length
and density correlations at $q \approx Q_p$ decaying as $\tau^{-2}$. 
Just like for the single dissipative quantum rotor, we expect this result, obtained in the 
large $N$ limit, to remain valid for $N = 2$. The form of the spatial correlations 
indicates that the system can be regarded as consisting of independent
 super-conducting ``puddles'' of size $\sim \xi$, each puddle behaving as a
 single ohmic quantum rotor.  In the TLL phase,  an analysis of the leading 
irrelevant operators shows that $\Phi(0,\tau) \sim \tau^{-2}$, at least. 
These results are in agreement with Griffiths' theorem~\cite{G67}.

We believe the model considered here to be  equivalent,  within the bosonization approach and for $\alpha \sim 1$, to the dissipative 2D XY model studied in~\cite{WTS05}. The  two dissipative phases found here correspond to those reported in~\cite{WTS05}. However, we note that on the line $\alpha = 0$ (an for $\alpha$ small too) the two models differ: whereas the model of~\cite{WTS05} undergoes a 
Kosterlitz-Thouless transition for large 
$g$ and $\alpha = 0$, and therefore, it is in a disordered (plasma) phase, our
model does not undergo such a transition and  instead exhibits a line of fixed
points for $\alpha = 0$ (the TLL phase). A detailed
comparison of the two models will be published elsewhere.

\emph{ Dual model:} A  model dual to the one discussed above can 
be realized if one considers an 1D metal of spin-$\frac{1}{2}$ fermions 
with attractive interactions (\emph{i.e.} a Luther-Emery liquid)
in front of a dirty metallic lead. Such a 1D metal
exhibits a spin gap~\cite{G04}, $\Delta_s$. For $T < \Delta_s$, 
single-electron hopping is suppressed and only hopping of pairs can
take place, which leads to a Josephson coupling to the lead:
\begin{equation}
H_J = - t_J  \int dx \: \left[ \Delta^{\dag}(x) \Delta_L(x) + {\rm h.c.} \right], 
\end{equation}
where $t_J \sim t^2/\Delta_s$, $t$ being the single-fermion hopping 
amplitude, and $\Delta_L(x) = \Psi_{\uparrow L}(x) 
\Psi_{\downarrow L }(x)$ and  $\Delta(x) = e^{i \theta(x)}$ the pairing
operators in the lead and the 1D metal, respectively. The field $\theta(x)$ 
is  dual  to the density field $\phi(x)$. Assuming that
$t_J$ is small, we integrate out the lead fermions  and  use that
$\langle \Delta_L(x,\tau) \Delta_L(0,0)\rangle_L \simeq 
|{\cal  G}^0_L(x,\tau)|^2 \: e^{ -|x|/\ell_p}$, where $\ell_p$
is the mean-free path and ${\cal G}^0_L(x,\tau) = [2\pi (v_{L F}\tau - ix)]^{-1}$
the single-particle Green's function of the lead electrons. Taking into account that 
$\theta(x,\tau)$  varies slowly on the scale of the correlation-length 
$\xi_s = \hbar v/\Delta_s \gg \ell_p \gg k^{-1}_{LF}$, where 
$k_{LF}$ is the Fermi momentum of the lead, we obtain 
\begin{eqnarray}
\tilde{S}_{\rm diss}[\theta] = -\frac{\eta_J}{\pi} \int dx 
d\tau d\tau' \,\frac{\cos\left[ \theta(x,\tau) - \theta(x,\tau') \right]}{|\tau-\tau'|^2}, 
\end{eqnarray}
for  $|\tau-\tau'| > \tau_c$, where $\tau_c \sim \tau_p = \ell_p/v_{LF}$ and
$\eta_J =  \nu_L  \tau_p (t_J/\hbar)^2/\hbar$, 
$\nu_L$ being the density of states of the lead. The RG flow
for this term can be obtained from Eqs.~(\ref{eq5},\ref{eq6}) by
replacing $g \to g^{-1}$ and $\alpha \to \alpha_J = (v \tau_c) \eta_J$,
and setting $p = \frac{1}{2}$ and $s = 1$. However, in a consistent
treatment~\cite{unpub}, the lead must be also treated as a source 
of dissipation for the density fluctuations.

{\emph{Conclusions:} We have analyzed a 1D metallic system with a few channels coupled to a metallic gate. In the absence of this coupling, the system is a
Tomonaga-Luttinger liquid (TLL). We have shown that this phase is stable if the
interactions in the wire are attractive or weakly repulsive, and the coupling
is small. For sufficient repulsion, the coupling to the gate
induces a phase transition to a gapless phase characterized by diffusive
charge excitations, in contrast to the acoustical plasmons found in one
dimensional conductors. For large compressibility, $v/g
\rightarrow 0$,  the coupling to the gate can induce a phase with a finite
spatial correlation length, and ohmic correlations in the temporal direction.

 Although for $g < g^*$ an arbitrarily small coupling to a gate destabilizes 
the TLL,  in practice finite temperature $T > 0$  or finite  length $L$ of the wire will cut-off the RG flow  before the system can exhibit the  full properties of the ordered (or disordered) phase described above. Thus, at finite
$T$ and $L$, it is important to find the optimal conditions for the coupling to
the gate to lead to measurable effects. 
The dimensionless quantity that measures the strength of the coupling to the gate is,
\emph{e.g.} for a 2D gate, $\alpha \approx  g (k_{GF}/k_F)^2 e^{-4p^2 k_F z_0}/\sigma_{2D}$, where $k_{GF}$ is the Fermi momentum of the gate electrons.
Therefore, the gate should be a high resistance metal, and the Fermi wavelength of the wire needs to be small compared to the distance to the gate. Note, however, that  close to the gate, the interactions within the wire strongly screened and $g \rightarrow 1$, and the TLL is stable. One  way around this problem would be
to use a granular gate, which provides ohmic dissipation but not metallic 
screening of the interactions for $q\approx 0$~\cite{G05}. It should be also mentioned that we have also neglected the possibility that disorder destabilizes the TLL before the effects of the gate are significant. These constraints suggest that a reasonable system where some of the phases discussed here can be observable is a weakly doped clean nanotube, where $k_{\rm F} z_0$ can be made small, coupled to a metallic gate with a short elastic mean free path or to a granular gate.

  We thank T. Giamarchi, A. Millis, A. Muramatsu, and P. Werner  for useful conversations.  This work has been supported by  \emph{Gipuzkoako Foru Aldundia} (MAC),   MEC (Spain) under Grants No  FIS2004-06490-C03-00 (MAC), MAT2002-0495-C02-01 (FG) and FIS2004-05120 (FS), and R. Areces Foundation (FS).   

\bibliography{disswire}
\end{document}